\documentclass{article}

\usepackage{mlspconf,amssymb,graphicx,booktabs,url,cite,multirow,todonotes}

\usepackage{amsmath}
\usepackage[amsmath,thmmarks]{ntheorem}
\usepackage{cleveref}

\usepackage[utf8]{inputenc}

\copyrightnotice{978-1-5090-6341-3/17/\$31.00 {\copyright}2017 IEEE}
\toappear{2017 IEEE International Workshop on Machine Learning for Signal Processing, Sept.\ 25--28, 2017, Tokyo, Japan}

\title{Joint Separation and Denoising of Noisy Multi-talker Speech using Recurrent Neural Networks and Permutation Invariant Training}

\name{Morten Kolbæk$^1$, Dong Yu$^2$, Zheng-Hua Tan$^1$, Jesper Jensen$^1$}
\address{$^1$Department of Electronic Systems, Aalborg University, Aalborg, Denmark\\
	$^2$Tencent AI Lab, Bellevue WA, USA.\\
	\{mok,zt,jje\}@es.aau.dk , dyu@tencent.com }

\begin{document}
\ninept

\maketitle
\begin{abstract}
In this paper we propose to use utterance-level Permutation Invariant Training\;(uPIT) for speaker independent multi-talker speech separation and denoising, simultaneously. 
Specifically, we train deep bi-directional Long Short-Term Memory\;(LSTM) Recurrent Neural Networks\;(RNNs) using uPIT, for single-channel speaker independent multi-talker speech separation in multiple noisy conditions, including both synthetic and real-life noise signals. 
We focus our experiments on generalizability and noise robustness of models that rely on various types of \emph{a priori} knowledge e.g. in terms of noise type and number of simultaneous speakers.

We show that deep bi-directional LSTM RNNs trained using uPIT in noisy environments can improve the Signal-to-Distortion Ratio\;(SDR) as well as the Extended Short-Time Objective Intelligibility\;(ESTOI) measure, on the speaker independent multi-talker speech separation and denoising task, for various noise types and Signal-to-Noise Ratios\;(SNRs). 
Specifically, we first show that LSTM RNNs can achieve large SDR and ESTOI improvements, when evaluated using known noise types, and that a single model is capable of handling multiple noise types with only a slight decrease in performance. 
Furthermore, we show that a single LSTM RNN can handle both two-speaker and three-speaker noisy mixtures, without \emph{a priori} knowledge about the exact number of speakers. Finally, we show that LSTM RNNs trained using uPIT generalize well to noise types not seen during training.

\end{abstract}

\section{Introduction}\label{sec:intro}
Focusing ones auditory attention towards a single speaker in a complex acoustic environment with multiple speakers and noise sources, is a task that  humans are extremely good at \cite{bronkhorst_cocktail_2000}. 
However, achieving similar performance with machines has so far not been possible \cite{mcdermott_cocktail_2009}, although it would be highly desirable for a vast range of applications, such as mobile communications, robotics, hearing aids, speaker verification systems, etc.   

Traditionally, speech denoising \cite{kolbaek_speech_2017,chen_long_2016,weninger_single-channel_2014,weninger_discriminatively_2014,erdogan_phase-sensitive_2015,chen_large-scale_2016} 
and multi-talker speech separation
\cite{du_speech_2014,yu_permutation_2017,hershey_deep_2016,isik_single-channel_2016,chen_deep_2017,weng_deep_2015,huang_joint_2015}  
have been considered as two separate tasks in the literature, although, for many applications both speech separation and denoising are desired. For example, in a human-machine interface the machine must be able to identify what is being said, and by who, before it can decide which signal to focus on, and consequently respond and act upon. 

The recent success of Deep Learning\cite{goodfellow_deep_2016} has revolutionized a large number of scientific fields, and is currently achieving state-of-the-art results on topics ranging from medical diagnosis \cite{gulshan_development_2016,esteva_dermatologist-level_2017} to Automatic Speech Recognition\;(ASR) \cite{xiong_achieving_2016,saon_english_2017}.  
Also the area of single-channel speech enhancement has seen improvement, with deep learning algorithms that have been reported to improve speech intelligibility for normal hearing, hearing impaired and cochlear implant users \cite{healy_algorithm_2015,chen_large-scale_2016,goehring_speech_2017,erdogan_deep_2017}. 
Speaker independent multi-talker speech separation, on the other hand, has so far not taken a similar leap forward, partly due to the long-lasting label permutation problem (further described in Section\;\ref{sec:train}), which has prevented progress on deep learning based techniques for this task.

Recently, two technical directions have been proposed for speaker independent multi-talker speech separation; a clustering based approach \cite{hershey_deep_2016,isik_single-channel_2016,chen_deep_2017}, and a regression based approach \cite{yu_permutation_2017,kolbaek_multi-talker_2017-1}. 
The clustering based approaches include the Deep Clustering\;(DPCL) techniques \cite{hershey_deep_2016,isik_single-channel_2016} and the DANet technique \cite{chen_deep_2017}.  
The regression based approaches include the Permutation Invariant Training\;(PIT) technique \cite{yu_permutation_2017} and the utterance-level PIT\;(uPIT) technique \cite{kolbaek_multi-talker_2017-1}.   
The general idea behind the DPCL and DANet techniques is that the mixture signal can be represented in an embedding space, e.g. using Recurrent Neural Networks\;(RNNs), where the different source signals in the mixture form clusters. These clusters are then identified using a clustering technique, such as K-means. The clustering based techniques have shown impressive performance on two-speaker and three-speaker mixtures. 
The regression based PIT and uPIT techniques, which are described in detail in Section\;\ref{sec:train} utilize a cost function that jointly optimizes the label assignment and regression error end-to-end, hence effectively solving the label permutation problem.

Both clustering based and regression based methods \cite{hershey_deep_2016,isik_single-channel_2016,chen_deep_2017,yu_permutation_2017,kolbaek_multi-talker_2017-1,erdogan_deep_2017} focus on ideal, noise-free training/testing conditions; i.e.\;situations where the mixtures contain clean speech only. For any practical application, background noise, e.g. due to interfering sound sources or non-ideal microphones must, be expected. However, it is yet to be known how these techniques perform, when tested in noisy conditions that reflect a realistic usage scenario. 

In this paper we apply the recently proposed uPIT technique \cite{kolbaek_multi-talker_2017-1} for speaker independent multi-talker speech separation and denoising, simultaneously. 
Specifically, we train deep bi-directional Long Short-Term Memory\;(LSTM) RNNs using uPIT for speaker independent multi-talker speech separation in multiple noisy conditions, including both synthetic and real-life, known and unknown, noise signals at various Signal-to-Noise Ratios\;(SNRs). 

To the authors knowledge, this is the first attempt to perform speech separation and denoising simultaneously in a deep learning framework; hence, no competing baseline has been identified for this particular task.

\section{Source Separation using Deep Learning}\label{sec:problem}

The goal of single-channel speech separation is to separate a mixture of multiple speakers into the individual speakers using a single microphone recording. Similarly, single-channel speech denoising aims to extract a single target speech signal from a noisy single channel recording. 

Let $x_s[n]$, $n = 1,2, \dots, N$, $s = 1,2, \dots, S$ be the time domain source signal of length $N$
from source $s$ and let the 
observed mixture signal be defined as 
\begin{equation}
	y[n] = \sum_{s=1}^S x_s[n],
\label{eqMixed}
\end{equation}
where $x_1[n]$ is a speech signal and $x_s[n]$, $s = 2, \dots , S$ can be either speech or additive noise signals. 
Furthermore, let ${X}_s(i,f)$ and ${Y}(i,f)$, $i = 1,\dots,K$, $f=1,\dots,L$ be the $L$-point Short-Time discrete Fourier Transforms (STFT) of $x_s[n]$ and $y[n]$, respectively.
Also, let $\mathbf{x}_{s,i} = \left[ {{X}_s(i,1)} ,\; {{X}_s(i,2)} ,\; \dots\; {{X}_s(i,\frac{L}{2}+1)} \right]^T \in \mathbb{C}^{\frac{L}{2}+1}$ and  $\mathbf{y}_{i} = \left[ {{Y}(i,1)} ,\; {{Y}(i,2)} ,\; \dots\; {{Y}(i,\frac{L}{2}+1)} \right]^T \in \mathbb{C}^{\frac{L}{2}+1}$ denote the single-sided STFT spectrum, at frame $i$, for sources $s=1, \dots ,S$ and the mixture signal, respectively. 

We define the magnitudes of the source signals and mixture signal as $A_s(i,f) \triangleq |{X}_s(i,f)|$ and $R(i,f)\triangleq |{Y}(i,f)|$, respectively, and their corresponding single-sided magnitude spectra as $\mathbf{a}_{s,i} = \left[ {{A_s}(i,1)} ,\; {{A_s}(i,2)} ,\; \dots\; {{A_s}(i,\frac{L}{2}+1)} \right]^T \in \mathbb{R}^{\frac{L}{2}+1}$ and $\mathbf{r}_{i} = \left[ {{R}(i,1)} ,\; {{R}(i,2)} ,\; \dots\; {{R}(i,\frac{L}{2}+1)} \right]^T \in \mathbb{R}^{\frac{L}{2}+1}$.
For separating the mixture signal $\mathbf{y}_{i}$ into estimated target signal magnitudes $\mathbf{a}_{s,i}$, $s=1, \dots ,S$, we adopt the approach from \cite{kolbaek_multi-talker_2017-1} and estimate a set of masks ${M_s}(t,f)$, $s=1, \dots ,S$ using bi-directional LSTM RNNs. 
Let $\mathbf{m}_{s,i} = \left[ {{M_s}(i,1)} \;,\; {{M_s}(i,2)} \;,\; \dots\; {{M_s}(i,\frac{L}{2}+1)} \right]^T \in \mathbb{R}^{\frac{L}{2}+1}$ be the ideal mask (to be defined in Sec.\;\ref{sec:mask}) for speaker $s$ at frame $i$.
The masks $\mathbf{{m}}_{s,i}$, $s=1, \dots ,S$ are then used to extract the target signal magnitudes as $\mathbf{{a}}_{s,i} = \mathbf{{m}}_{s,i} \circ \mathbf{r}_i$, $s=1, \dots ,S$, $i=1, \dots ,K$ where $\circ$ is the element-wise product, i.e. the Hadamard product. 
Similarly, when the masks are estimated by a deep learning model we arrive at the estimated signal magnitudes as $\mathbf{\hat{a}}_{s,i} = \mathbf{\hat{m}}_{s,i} \circ \mathbf{r}_i$, $s=1, \dots ,S$, $i=1, \dots ,K$.
The overlap-and-add technique and the inverse discrete Fourier transform, using the phase of the mixture signal, is used for reconstructing $\mathbf{\hat{a}}_{s,i}$, $i=1, \dots ,K$ in the time domain.

\subsection{Mask Estimation and Loss functions}\label{sec:mask}
A large number of training targets and loss functions have been proposed for masking based source separation \cite{erdogan_phase-sensitive_2015,wang_training_2014,erdogan_deep_2017}. Since the one reasonable goal is to have an accurate reconstruction, a loss function based on the reconstruction error instead of the mask estimation error is preferable \cite{erdogan_deep_2017}.  

In \cite{kolbaek_multi-talker_2017-1}, different such loss functions were investigated for speaker independent multi-talker speech separation and the best performing one was found to be the Phase-Sensitive Approximation\;(PSA) loss function \cite{erdogan_phase-sensitive_2015}, which for frame $i$ is given as  
\begin{equation}
\begin{aligned}
J_{i}^{PSA} = \;
& \sum_{s=1}^S \| \mathbf{\hat{m}}_{s,i}\circ\mathbf{{r}}_{i} - \mathbf{a}_{s,i}\cos(\mathbf{\phi}_{s,i}) \|_2^2  \\
=\;& \sum_{s=1}^S \| \mathbf{\hat{a}}_{s,i} - \mathbf{a}_{s,i}\cos(\mathbf{\phi}_{s,i}) \|_2^2  \\
\end{aligned}
\label{eqLoss}
\end{equation}
where $\mathbf{\phi}_{s,i} = \mathbf{\phi}_{y,i} - \mathbf{\phi}_{s,i} $ is the element-wise phase difference between the mixture $\mathbf{y}_i$ and the source $\mathbf{x}_{s,i}$ and $||\cdot||_2$ is the $\ell^2$-norm.   

In contrast to the classical squared error loss function, i.e. Eq.\;\eqref{eqLoss} without the cosine term, the PSA loss function accounts for some of the errors introduced by the noisy phase used in the reconstruction. 
When the PSA loss function is used for mask estimation, the actual mask estimated is the Ideal Phase-Sensitive Filter\;(IPSF) \cite{erdogan_phase-sensitive_2015}, which due to the phase correction property is preferable over other commonly used masks such as the Ideal Ratio Mask, or the Ideal Amplitude Mask \cite{erdogan_deep_2017}.

\section{Permutation invariant training}\label{sec:train}
Permutation Invariant Training\;(PIT) is a generalization of the traditional approach for training Deep Neural Networks\;(DNNs) for regression based source separation problems, such as speaker separation or denoising. 

For training a DNN based source separation model with $S$ output masks, $\mathbf{\hat{m}}_{s,i}$, $s=1, \dots ,S$, an MSE criterion is typically used and is computed between the true sources $\mathbf{a}_{s,i}$ and the estimated sources $\mathbf{\hat{a}}_{s,i} = \mathbf{\hat{m}}_{s,i} \circ \mathbf{r}_i$, $s = 1, \dots , S$, $i=1,\dots,K$. However, with multiple outputs, it is not trivial to pair the outputs with the correct targets.   
The commonly used approach for pairing a given output $\mathbf{\hat{a}}_{s,i}$ to a certain target $\mathbf{a}_{s,i}$ is to predefine the targets into an ordered list, such that output one is always paired with e.g. target one, i.e. $(\mathbf{a}_{1,i},\mathbf{\hat{a}}_{1,i})$, output two with target two $(\mathbf{a}_{2,i},\mathbf{\hat{a}}_{2,i})$, etc.

For tasks such as speech denoising with a single speaker in noise, or speech separation of known speakers\cite{huang_joint_2015}, simply predefining the ordering of the targets works well and the DNN can learn to correctly separate the sources and will provide the correct source at the output corresponding to the correct target. 
However, for mixtures containing similar signals, such as unknown equal energy male speakers, this standard training approach fails to converge \cite{yu_permutation_2017,weng_deep_2015,hershey_deep_2016}. 
Empirically, it is found that DNNs are likely to change permutation from one frame to another for highly similar sources. Hence, predefining the ordering of the targets, might not be the optimal solution, and clearly a bad solution for certain types of signals. This phenomenon, and the challenge of choosing the output-target permutation during training, is commonly known as the label permutation or ambiguity problem \cite{weng_deep_2015,hershey_deep_2016,chen_deep_2017,kolbaek_multi-talker_2017-1}. 

In \cite{yu_permutation_2017} a solution to the label permutation problem was proposed, where targets are provided as a set instead of an ordered list and the output-target permutation $\theta$, for a given frame, is defined as the permutation that minimizes the cost function in question (e.g. squared error) over all possible permutations $\mathcal{P}$. Following this approach combined with the PSA loss function, a permutation invariant training criterion and corresponding error $J_i^{PIT}$, for the $i^{th}$ frame, can be formulated as
\begin{equation}
\begin{aligned}
J_i^{PIT} = \;&  \underset{\theta\in \mathcal{P}}{\text{min}} 
& \sum_{s=1}^S \| \mathbf{\hat{a}}_{s,i} - \mathbf{a}_{\theta(s),i} \cos(\mathbf{\phi}_{s,i}) \|_2^2.  \\
\end{aligned}
\label{eqPITseg}
\end{equation}

As shown in \cite{yu_permutation_2017}, Eq.\;\eqref{eqPITseg} effectively solves the label permutation problem. However, since PIT as defined in Eq.\;\eqref{eqPITseg} operates on frames, the DNN only learns to separate the input mixtures into sources at the frame level, and not the utterance level. 
In practice, this means that the mixture might be correctly separated, but the frames belonging to a particular speaker are not assigned the same output index throughout the utterance and without exact knowledge about the speaker-output permutation, it is very difficult to correctly reconstruct the separated sources.
In order to have the sources separated at the utterance-level, so that all frames from a particular output belong to the same source, additional speaker tracing or very large input-output contexts are needed \cite{yu_permutation_2017}.

\subsection{Utterence-level Permutation Invariant Training } \label{subsec:pittrace}
In \cite{kolbaek_multi-talker_2017-1} an extension to PIT, known as utterance-level PIT\;(uPIT) was proposed for solving the speaker-output permutation problem. In uPIT, the output-target permutation $\theta$ is given as the permutation that gives the minimum squared error over all possible permutations for the entire utterance, instead of only a single frame. Formally, the utterance-level permutation used for training is found as   
\begin{equation}
\begin{aligned}
\theta^\ast = \underset{\theta\in \mathcal{P}}{\text{argmin}} \sum_{s=1}^S \sum_{i=1}^K  \| \hat{\mathbf{a}}_{s,i} -  \mathbf{a}_{\theta(s),i}\cos(\mathbf{\phi_{\theta(s),i}}) \|_2^2  
\end{aligned}
\label{eqPITutt}
\end{equation}
and the permutation $\theta^\ast$ is then used \emph{for all} frames within the current utterance, hence an utterance-level loss $J_{\theta^\ast,i}^{uPIT}$ for the $i^{th}$ frame in a given utterance is defined as
\begin{equation}
\begin{aligned}
J_{\theta^\ast,i}^{uPIT} = \;& \sum_{s=1}^S \| \mathbf{\hat{a}}_{s,i} - \mathbf{a}_{\theta^\ast(s),i} \cos(\mathbf{\phi_{\theta^\ast(s),i}}) \|_2^2.  \\
\end{aligned}
\label{eqPITutt1}
\end{equation}
Using the same permutation \emph{for all} frames in the entire utterance has the consequence that the smallest per-frame error will not always be used for training as with original PIT. Instead the smallest \emph{per-utterance} error will be used, which enforces the estimated sources to stay at the same DNN outputs for the entire utterance. Ideally, this means that each DNN output contains a single source. Finally, since the whole utterance is needed for computing the utterance-level permutation in Eq.\;\eqref{eqPITutt}, RNNs are a natural choice of DNN model for this loss function.

\section{Experimental Design}\label{sec:expsetup}
To study the noise robustness of the uPIT technique, we have conducted several experiments with noise corrupted mixtures of multiple speakers. 
Since uPIT uses the noise-free source signals as training targets, a denoising capability is already present in the uPIT framework. By simply adding noise to the multi-speaker input mixture, a model trained with uPIT will not only learn to separate the sources but also to remove the noise.

\subsection{Noise-free Multi-talker Speech Mixtures} \label{subsec:speechdata}
We have used the noise-free two-speaker mixture (WSJ0-2mix) and three-speaker mixture (WSJ0-3mix)\footnote{Available at: http://www.merl.com/demos/deep-clustering} datasets for all experiments conducted in this paper. These datasets have been used in \cite{hershey_deep_2016,kolbaek_multi-talker_2017-1,yu_permutation_2017,isik_single-channel_2016}, which allows us to relate the performance of uPIT in noisy conditions with the performance in noise-free conditions. 
The feature representation is based on 129-dimensional STFT magnitude spectra, extracted from a 256 point STFT using a sampling frequency of 8 kHz, a hanning window size of 32 ms and a 16 ms frame shift.

The WSJ0-2mix dataset was derived from the WSJ0 corpus \cite{garofolo_csr-i_1993}. 
The WSJ0-2mix training set and validation set contain two-speaker mixtures generated by randomly selecting pairs of utterances from 49 male and 51 female speakers from the WSJ0 training set entitled si\_tr\_s. 
The two utterances are then mixed with a difference in active speech level \cite{noauthor_itu_1993} uniformly chosen between 0\;dB and 5\;dB. 
The training and validation sets consist of 20000 and 5000 mixtures, respectively, which is equivalent to approximately 30 hours of training data and 5 hours of validation data. The test set was similarly generated using utterances from 16 speakers from the WSJ0 validation set si\_dt\_05 and evaluation set si\_et\_05, and consists of 5000 mixtures or approximately 5 hours of data. That is, the speakers in the test set are different from the speakers in the training and validation sets. The WSJ0-3mix dataset was generated using a similar approach but contains mixtures of speech from three speakers.

Since we want a single RNN architecture that can handle both two-speaker and three-speaker mixtures, we have chosen a model architecture with three outputs. The specific architecture is described in detail in Sec.\;\ref{subsec:modelarch}. To ensure that the model can handle both two-speaker and three-speaker mixtures, the model must be trained on both scenarios, so we have combined the WSJ0-2mix and WSJ0-3mix datasets into a larger WSJ0-2+3mix dataset. To allow this fusion, we have extended the WSJ0-2mix dataset with a third "silent" speaker, such that the combined WSJ0-2+3mix dataset consists of only three speaker mixtures, but half of the mixtures contain three speakers, and the remaining half contain two speaker mixtures (and a "silent speaker"). To minimize the risk of numerical issues, e.g. in computing ideal masks, the third "silent" speaker consists of white Gaussian noise with an average energy level 70 dB below the average energy of the other two speakers in the mixture.  

\subsection{Noisy Multi-talker Speech Mixtures} \label{subsec:noisedata}
To simulate noisy environments, we follow the common approach \cite{kolbaek_speech_2017} for generating noisy mixtures with additive noise and simply add the noise-free WSJ0-2+3mix mixture signal with a noise signal. To achieve a certain SNR the noise signal is scaled based on the active speech level of the noise-free mixture signal as per ITU P.56 \cite{noauthor_itu_1993}. 

To evaluate the robustness of the uPIT model against a stationary noise type, we use a synthetic Speech Shaped Noise\;(SSN) signal.  
The SSN noise signal is constructed by filtering a Gaussian white noise sequence through a $12^{th}$-order all-pole filter with coefficients found from Linear Predictive Coding\;(LPC) analysis of 100 randomly chosen TIMIT sentences \cite{garofolo_darpa_1993}.  

To evaluate the robustness against a highly non-stationary noise type we use a synthetic 6-speaker Babble\;(BBL) noise. 
The BBL noise signal is also based on TIMIT. The corpus, which consists of a total of 6300 spoken sentences, is randomly divided  into 6 groups of 1050 concatenated utterances. Each group is then normalized to unit energy and truncated to equal length followed by  addition of the six groups. This results in a BBL noise sequence with a duration of over 50 min. 

To evaluate the robustness against realistic noise types we use the street\;(STR), cafeteria\;(CAF), bus\;(BUS), and pedestrian\;(PED) noise signals from the CHiME3 dataset\cite{barker_third_2015}. These noise signals are real-life recordings in their respective environments.   

All six noise signals are divided into a 40 min.\;training sequence, a 5 min.\;validation sequence and a 5 min.\;test sequence. That is, the noise signals used for training and validation are different from the sequence used for testing.

\subsection{Model Architectures and Training} \label{subsec:modelarch}
For evaluating uPIT in noisy environments we have trained a total of seven bi-directional LSTM RNNs \cite{hochreiter_long_1997}, using the training conditions, i.e. datasets and noise types, presented in Table.\;\ref{tab:models}. 
\begin{table}
	\caption{Training conditions for different models.}
	\label{tab:models}
	\centering
	\setlength\tabcolsep{5pt} 
	\resizebox{0.8\columnwidth}{!}{
	\begin{tabular}{c|cc}
		\toprule
		Model ID & \multicolumn{2}{|l} {Dataset + Noise type (SNR: -5 dB -- 10 dB)}  \\ 
		\midrule
		LSTM1  & \multicolumn{2}{|l} {WSJ0-2+3mix + SSN } 	       \\ 
		LSTM2   & \multicolumn{2}{|l} {WSJ0-2+3mix + BBL }  	  \\ 
		LSTM3   & \multicolumn{2}{|l} {WSJ0-2+3mix + STR }   	  \\ 
		LSTM4  & \multicolumn{2}{|l} {WSJ0-2+3mix + CAF } 	    \\ 
		LSTM5  &\multicolumn{2}{|l} {WSJ0-2+3mix + SSN + BBL + STR + CAF }    	  \\ 
		LSTM6  & \multicolumn{2}{|l} {WSJ0-2mix + BBL }   	  \\ 
		LSTM7  & \multicolumn{2}{|l} {WSJ0-3mix + BBL } \\ \bottomrule
	\end{tabular}}
\end{table}
LSTM1-5 were trained on the WSJ0-2+3mix dataset, which contains a mix of both two-speaker and three-speaker mixtures. LSTM1-4 are noise type specific in the sense that they were trained using only a single noise type. LSTM5 was trained on all four noise types. LSTM6 and LSTM7 were trained using WSJ0-2mix and WSJ0-3mix datasets, respectively, and only a single noise type. 
LSTM5 will show the performance degradation, if any, when less \emph{a priori} knowledge about the noise types is available. Similarly, LSTM6-7 will show the potential performance improvement if the number of speakers in the mixture is known \emph{a priori}. 
Each mixture in the dataset was corrupted with noise at a specific SNR, uniformly chosen between -5\;dB and 10\;dB.   

Each model has three bi-directional LSTM layers, and a fully-connected output layer with ReLU \cite{goodfellow_deep_2016} activation functions. LSTM1-5 and LSTM7 have 1280 LSTM cells in each layer and LSTM6 has 896 cells, to be compliant with \cite{kolbaek_multi-talker_2017-1}. 
The input dimension is 129, i.e a single frame $\mathbf{r}_i$ and the output dimension is $3\times129 = 387$, i.e. $\mathbf{\hat{a}}_{s,i}$, $s=1,2,3$. We apply 50\% dropout \cite{goodfellow_deep_2016} between the LSTM layers, and the outputs from the forward and backward LSTMs, from one layer, are concatenated before they are used as input to the subsequent layer. 
LSTM6 has approximately $46\cdot10^6$ trainable parameters, and LSTM1-5 and 7 have approximately $94\cdot 10^6$ trainable parameters, which are found using stochastic gradient descent with gradients found by backpropagation.  
In all the experiments, the maximum number of epochs was set to 200 and the learning rates were set to $2 \cdot 10^{-5}$ per sample initially, and scaled down by $0.7$ when the training cost increased on the training set. The training was terminated when the learning rate got below $10^{-10}$. Each minibatch contains 8 randomly selected utterances.
All models are implemented using the Microsoft Cognitive Toolkit (CNTK) \cite{agarwal_introduction_2014}\footnote{Available at: https://www.cntk.ai/}.

\section{Experimental Results}\label{sec:exp}
We evaluated the noise robustness of LSTM1-7 using the Signal to Distortion Ratio\;(SDR) \cite{vincent_performance_2006} and the Extended Short-Time Objective Intelligibility\;(ESTOI) measure \cite{jensen_algorithm_2016}.
The SDR is an often used performance metric for source separation and is defined in dB. 
The ESTOI measure estimates speech intelligibility and has been found to be highly correlated with human listening tests \cite{jensen_algorithm_2016}, especially for modulated maskers. The ESTOI measure is defined in the range $[-1, 1]$, and higher is better.   
When evaluating SDR and ESTOI, we choose the output-target permutation that maximizes the given performance metric. Furthermore, when evaluating two-speaker mixtures, we identify the silent speaker as the output with the least energy and then compute the performance metric based on the remaining two outputs.

\Cref{tab:sdr_ssn,tab:sdr_bbl,tab:sdr_str,tab:sdr_caf} summarize the SDR \emph{improvements} achieved by LSTM1-5 on two and three-speaker mixtures corrupted by SSN, BBL, STR, and CAF noise, respectively. The improvements are relative to the SDR of the noisy mixture without processing ("No Proc." in Tables). 
\Cref{tab:stoi_ssn,tab:stoi_bbl,tab:stoi_str,tab:stoi_caf} summarize ESTOI \emph{improvements} achieved by the same models in similar conditions. 
We evaluate the models at the challenging SNR of $-5$\;dB, as well as at $0$,\; $5$\;, and $20$\;dB. At an input SNR of $-5$\;dB, speech intelligibility, as estimated by ESTOI, is severely degraded, primarily due to the noise component, whereas speech intelligibility degradation at $20$\;dB is primarily caused by the competing talkers in the mixture itself.
As a reference, we also reported the IPSF performance, which uses oracle information and therefore serves as an upper performance bound on this particular task.

From \Cref{tab:sdr_ssn,tab:sdr_bbl,tab:sdr_str,tab:sdr_caf,tab:stoi_ssn,tab:stoi_bbl,tab:stoi_str,tab:stoi_caf} we see that all noise-type specific models, i.e. LSTM1-4, in general achieve large SDR and ESTOI improvements with an average improvement of $9.1$\;dB and $0.18$ for SDR and ESTOI, respectively, for two-speaker mixtures and $7.2$\;dB and $0.13$, respectively, for three-speaker mixtures. 
Furthermore, we see that LSTM5 performs only slightly worse than the noise type specific models, which is interesting, since LSTM5 and LSMT1-4 have all been trained with 60 hours of speech, but LSTM5 have only seen $15$ hours of each noise type, compared to $60$ hours for LSTM1-4. 
We also observe that the highly non-stationary BBL noise seems to be considerably harder than the three other noise types, which corresponds well with existing literature \cite{kolbaek_speech_2017,loizou_speech_2013,erkelens_minimum_2007}.       

\Cref{tab:sdr_bbl_spkr,tab:stoi_bbl_spkr} summarize the performance of LSTM6 and LSTM7. We observe that both models perform approximately similar to the noise-type-general LSTM5. More surprisingly, we see that LSTM2 consistently outperforms both LSTM6 and LSTM7, which corresponds well with a similar observation in the noise-free case in \cite{kolbaek_multi-talker_2017-1}. 
These results are of great importance, since they show that training a model on noisy three-speaker mixtures helps the model separating noisy two-speaker mixtures, and vice versa.

\Cref{tab:sdr_ped,tab:stoi_ped} summarize the performance of LSTM5, when evaluated using speech mixtures corrupted with the two unknown noise types, PED and BUS, i.e. noise types not included in the training set. The score with respect to the BUS noise type is on the left of the vertical bar ($|$) and PED is on the right. We see that LSTM5 achieves large SDR and ESTOI improvements for both noise types, at almost all SNRs. More importantly, we observe that the scores are comparable with, and in some cases even exceed, the performance of LSTM5, when it was evaluated using known noise types as reported in \Cref{tab:sdr_ssn,tab:sdr_bbl,tab:sdr_str,tab:sdr_caf,tab:stoi_ssn,tab:stoi_bbl,tab:stoi_str,tab:stoi_caf}. These results indicate that LSTM5 is relatively robust against variations in the noise distribution.

In general, we observe SDR improvements for all models that are comparable in magnitude with the noise-free case \cite{yu_permutation_2017,kolbaek_multi-talker_2017-1,hershey_deep_2016,isik_single-channel_2016}. 
However, the SDR measure, as well as ESTOI, do not differentiate between distortions from other speakers (such as Source to Inference Ratio from \cite{vincent_performance_2006}) and distortion from the noise source. This means that the trade-off between speech separation and noise-reduction is yet to be fully understood. We leave this topic for future research.

\begin{table}
	\caption{SDR improvements for LSTM1 and 5 tested on SSN.}
	\label{tab:sdr_ssn}
	\centering
	\setlength\tabcolsep{5pt} 
	\resizebox{1.0\columnwidth}{!}{%
		\begin{tabular}{c|cccc|cccc}
			\toprule
			& \multicolumn{4}{|c|} {2-Speaker } & \multicolumn{4}{c} {3-Speaker} \\ \midrule
			\begin{tabular}[c]{@{}c@{}}SNR \\ {[dB]}\end{tabular} 	& 
			\begin{tabular}[c]{@{}c@{}}No \\ Proc.\end{tabular} 	& 
			\begin{tabular}[c]{@{}c@{}} IPSF \end{tabular} 			& 
			\begin{tabular}[c]{@{}c@{}}LSTM1\end{tabular} 		& 
			\begin{tabular}[c]{@{}c@{}}LSTM5\end{tabular} 		&   
			\begin{tabular}[c]{@{}c@{}}No \\ Proc.\end{tabular} 	& 
			\begin{tabular}[c]{@{}c@{}} IPSF \end{tabular} 			& 
			\begin{tabular}[c]{@{}c@{}}LSTM1\end{tabular} 		& 
			\begin{tabular}[c]{@{}c@{}}LSTM5\end{tabular} \\ 
			\midrule
						-5 & -8.8 & 15.9 & 9.6 & 9.4 & -10.3 & 16.6 & 8.0 & 7.8 \\ 
			0 & -5.1 & 14.5 & 9.1 & 9.0 & -7.0 & 15.2 & 7.6 & 7.4 \\ 
			5 & -2.4 & 13.9 & 8.6 & 8.4 & -4.8 & 14.6 & 7.0 & 6.9 \\ 
			20 & 0.0 & 14.8 & 8.7 & 8.8 & -3.0 & 15.1 & 6.6 & 6.7 \\ \midrule
			Avg. & -4.1 & 14.8 & 9.0 & 8.9 & -6.3 & 15.4 & 7.3 & 7.2 \\ \bottomrule
		\end{tabular}}
	%
	\caption{SDR improvements for LSTM2 and 5 tested on BBL.}
	\label{tab:sdr_bbl}
	\centering
	\setlength\tabcolsep{5pt} 
	\resizebox{1.0\columnwidth}{!}{%
		\begin{tabular}{c|cccc|cccc}
			\toprule
			& \multicolumn{4}{|c|} {2-Speaker } & \multicolumn{4}{c} {3-Speaker} \\ \midrule
			\begin{tabular}[c]{@{}c@{}}SNR \\ {[dB]}\end{tabular} 	& 
			\begin{tabular}[c]{@{}c@{}}No \\ Proc.\end{tabular} 	& 
			\begin{tabular}[c]{@{}c@{}} IPSF \end{tabular} 			& 
			\begin{tabular}[c]{@{}c@{}}LSTM2\end{tabular} 		& 
			\begin{tabular}[c]{@{}c@{}}LSTM5\end{tabular} 		&   
			\begin{tabular}[c]{@{}c@{}}No \\ Proc.\end{tabular} 	& 
			\begin{tabular}[c]{@{}c@{}} IPSF \end{tabular} 			& 
			\begin{tabular}[c]{@{}c@{}}LSTM2\end{tabular} 		& 
			\begin{tabular}[c]{@{}c@{}}LSTM5\end{tabular} \\ 
			\midrule
						-5 & -8.9 & 17.2 & 6.0 & 5.4 & -10.4 & 17.8 & 4.4 & 3.8 \\ 
			0 & -5.1 & 15.4 & 8.1 & 7.6 & -7.1 & 16.0 & 6.3 & 5.8 \\ 
			5 & -2.4 & 14.5 & 8.5 & 8.1 & -4.8 & 15.1 & 6.7 & 6.5 \\ 
			20 & 0.0 & 14.8 & 9.0 & 8.8 & -3.0 & 15.2 & 6.8 & 6.7 \\ \midrule
			Avg. & -4.1 & 15.5 & 7.9 & 7.5 & -6.3 & 16.0 & 6.0 & 5.7 \\ \bottomrule
		\end{tabular}}
	%
	\caption{SDR improvements for LSTM3 and 5 tested on STR.}
	\label{tab:sdr_str}
	\centering
	\setlength\tabcolsep{5pt} 
	\resizebox{1.0\columnwidth}{!}{%
		\begin{tabular}{c|cccc|cccc}
			\toprule
			& \multicolumn{4}{|c|} {2-Speaker } & \multicolumn{4}{c} {3-Speaker} \\ \midrule
			\begin{tabular}[c]{@{}c@{}}SNR \\ {[dB]}\end{tabular} 	& 
			\begin{tabular}[c]{@{}c@{}}No \\ Proc.\end{tabular} 	& 
			\begin{tabular}[c]{@{}c@{}} IPSF \end{tabular} 			& 
			\begin{tabular}[c]{@{}c@{}}LSTM3\end{tabular} 		& 
			\begin{tabular}[c]{@{}c@{}}LSTM5\end{tabular} 		&   
			\begin{tabular}[c]{@{}c@{}}No \\ Proc.\end{tabular} 	& 
			\begin{tabular}[c]{@{}c@{}} IPSF \end{tabular} 			& 
			\begin{tabular}[c]{@{}c@{}}LSTM3\end{tabular} 		& 
			\begin{tabular}[c]{@{}c@{}}LSTM5\end{tabular} \\ 
			\midrule
						-5 & -8.9 & 18.2 & 11.5 & 11.5 & -10.4 & 18.6 & 9.7 & 9.6 \\ 
			0 & -5.2 & 16.2 & 10.2 & 10.2 & -7.1 & 16.7 & 8.4 & 8.3 \\ 
			5 & -2.4 & 14.9 & 9.2 & 9.1 & -4.8 & 15.5 & 7.3 & 7.2 \\ 
			20 & 0.0 & 14.9 & 8.9 & 8.8 & -3.0 & 15.2 & 6.6 & 6.7 \\ \midrule
			Avg. & -4.1 & 16.1 & 9.9 & 9.9 & -6.3 & 16.5 & 8.0 & 7.9 \\ \bottomrule
		\end{tabular}}
	%
	\caption{SDR improvements for LSTM4 and 5 tested on CAF.}
	\label{tab:sdr_caf}
	\centering
	\setlength\tabcolsep{5pt} 
	\resizebox{1.0\columnwidth}{!}{%
		\begin{tabular}{c|cccc|cccc}
			\toprule
			& \multicolumn{4}{|c|} {2-Speaker } & \multicolumn{4}{c} {3-Speaker} \\ \midrule
			\begin{tabular}[c]{@{}c@{}}SNR \\ {[dB]}\end{tabular} 	& 
			\begin{tabular}[c]{@{}c@{}}No \\ Proc.\end{tabular} 	& 
			\begin{tabular}[c]{@{}c@{}} IPSF \end{tabular} 			& 
			\begin{tabular}[c]{@{}c@{}}LSTM4\end{tabular} 		& 
			\begin{tabular}[c]{@{}c@{}}LSTM5\end{tabular} 		&   
			\begin{tabular}[c]{@{}c@{}}No \\ Proc.\end{tabular} 	& 
			\begin{tabular}[c]{@{}c@{}} IPSF \end{tabular} 			& 
			\begin{tabular}[c]{@{}c@{}}LSTM4\end{tabular} 		& 
			\begin{tabular}[c]{@{}c@{}}LSTM5\end{tabular} \\ 
			\midrule
						-5 & -8.9 & 18.2 & 10.0 & 9.9 & -10.4 & 18.6 & 8.4 & 8.2 \\ 
			0 & -5.1 & 16.3 & 9.7 & 9.5 & -7.1 & 16.8 & 7.9 & 7.7 \\ 
			5 & -2.4 & 15.1 & 9.0 & 8.9 & -4.8 & 15.6 & 7.1 & 6.9 \\ 
			20 & 0.0 & 14.8 & 8.8 & 8.8 & -3.0 & 15.2 & 6.7 & 6.6 \\ \midrule
			Avg. & -4.1 & 16.1 & 9.4 & 9.3 & -6.3 & 16.6 & 7.5 & 7.3 \\ \bottomrule
		\end{tabular}}
	%
	\caption{SDR improvements for LSTM6, 7 and 5 tested on BBL.}
	\label{tab:sdr_bbl_spkr}
	\centering
	\setlength\tabcolsep{5pt} 
	\resizebox{1.0\columnwidth}{!}{%
		\begin{tabular}{c|cccc|cccc}
			\toprule
			& \multicolumn{4}{|c|} {2-Speaker } & \multicolumn{4}{c} {3-Speaker} \\ \midrule
			\begin{tabular}[c]{@{}c@{}}SNR \\ {[dB]}\end{tabular} 	& 
			\begin{tabular}[c]{@{}c@{}}No \\ Proc.\end{tabular} 	& 
			\begin{tabular}[c]{@{}c@{}} IPSF \end{tabular} 			& 
			\begin{tabular}[c]{@{}c@{}}LSTM6\end{tabular} 		& 
			\begin{tabular}[c]{@{}c@{}}LSTM5\end{tabular} 		&   
			\begin{tabular}[c]{@{}c@{}}No \\ Proc.\end{tabular} 	& 
			\begin{tabular}[c]{@{}c@{}} IPSF \end{tabular} 			& 
			\begin{tabular}[c]{@{}c@{}}LSTM7\end{tabular} 		& 
			\begin{tabular}[c]{@{}c@{}}LSTM5\end{tabular} \\ 
			\midrule
		    			-5 & -8.9 & 17.2 & 5.6 & 5.4 & -10.4 & 17.8 & 4.0 & 3.8 \\ 
			0 & -5.1 & 15.4 & 7.7 & 7.6 & -7.1 & 16.0 & 5.7 & 5.8 \\ 
			5 & -2.4 & 14.5 & 8.0 & 8.1 & -4.8 & 15.1 & 6.3 & 6.5 \\ 
			20 & 0.0 & 14.9 & 8.4 & 8.8 & -3.0 & 15.2 & 6.4 & 6.7 \\ \midrule
			Avg. & -4.1 & 15.5 & 7.4 & 7.5 & -6.3 & 16.0 & 5.6 & 5.7 \\ \bottomrule
		\end{tabular}}
	%
    \caption{SDR improvements for LSTM5 tested on BUS $|$ PED.}
	\label{tab:sdr_ped}
	\centering
	\setlength\tabcolsep{5pt} 
	\resizebox{0.96\columnwidth}{!}{%
		\begin{tabular}{c|ccc|ccc}
			\toprule
			& \multicolumn{3}{|c|} {2-Speaker } & \multicolumn{3}{c} {3-Speaker} \\ \midrule
			\begin{tabular}[c]{@{}c@{}}SNR \\ {[dB]}\end{tabular} 	& 
			\begin{tabular}[c]{@{}c@{}}No \\ Proc.\end{tabular} 	& 
			\begin{tabular}[c]{@{}c@{}} IPSF \end{tabular} 			& 
			\begin{tabular}[c]{@{}c@{}}LSTM5\end{tabular} 		&   
			\begin{tabular}[c]{@{}c@{}}No \\ Proc.\end{tabular} 	& 
			\begin{tabular}[c]{@{}c@{}} IPSF \end{tabular} 			& 
			\begin{tabular}[c]{@{}c@{}}LSTM5\end{tabular} \\ 
			\midrule
						-5 & -9.0 $|$ -8.9 & 19.6 $|$ 16.7 & 11.7 $|$ 7.3 & -10.5 $|$ -10.4 & 19.9 $|$ 17.4 & 9.7 $|$ 5.7 \\ 
			0 & -5.2 $|$ -5.2 & 17.3 $|$ 14.9 & 10.7 $|$ 7.8 & -7.2 $|$ -7.1 & 17.6 $|$ 15.7 & 8.5 $|$ 6.3 \\ 
			5 & -2.4 $|$ -2.4 & 15.7 $|$ 14.1 & 9.5 $|$ 7.9 & -4.8 $|$ -4.8 & 16.1 $|$ 14.8 & 7.4 $|$ 6.3 \\ 
			20 & 0.0 $|$ 0.0 & 14.9 $|$ 14.8 & 8.8 $|$ 8.7 & -3.0 $|$ -3.0 & 15.2 $|$ 15.2 & 6.7 $|$ 6.7 \\ \midrule
			Avg. & -4.1 $|$ -4.1 & 16.9 $|$ 15.1 & 10.2 $|$ 7.9 & -6.4 $|$ -6.3 & 17.2 $|$ 15.8 & 8.1 $|$ 6.2 \\ \bottomrule
		\end{tabular}}
\end{table}

\begin{table}
	\caption{ESTOI improvements for LSTM1 and 5 tested on SSN.}
	\label{tab:stoi_ssn}
	\centering
	\setlength\tabcolsep{5pt} 
	\resizebox{1.0\columnwidth}{!}{%
		\begin{tabular}{c|cccc|cccc}
			\toprule
			& \multicolumn{4}{|c|} {2-Speaker } & \multicolumn{4}{c} {3-Speaker} \\ \midrule
			\begin{tabular}[c]{@{}c@{}}SNR \\ {[dB]}\end{tabular} 	& 
			\begin{tabular}[c]{@{}c@{}}No \\ Proc.\end{tabular} 	& 
			\begin{tabular}[c]{@{}c@{}} IPSF \end{tabular} 			& 
			\begin{tabular}[c]{@{}c@{}}LSTM1\end{tabular} 		& 
			\begin{tabular}[c]{@{}c@{}}LSTM5\end{tabular} 		&   
			\begin{tabular}[c]{@{}c@{}}No \\ Proc.\end{tabular} 	& 
			\begin{tabular}[c]{@{}c@{}} IPSF \end{tabular} 			& 
			\begin{tabular}[c]{@{}c@{}}LSTM1\end{tabular} 		& 
			\begin{tabular}[c]{@{}c@{}}LSTM5\end{tabular} \\ 
			\midrule
						-5 & 0.18 & 0.65 & 0.17 & 0.16 & 0.14 & 0.69 & 0.10 & 0.09 \\ 
			0 & 0.29 & 0.58 & 0.23 & 0.22 & 0.22 & 0.63 & 0.15 & 0.14 \\ 
			5 & 0.39 & 0.50 & 0.23 & 0.22 & 0.29 & 0.58 & 0.17 & 0.16 \\ 
			20 & 0.54 & 0.39 & 0.17 & 0.18 & 0.38 & 0.53 & 0.15 & 0.15 \\ \midrule
			Avg. & 0.35 & 0.53 & 0.20 & 0.20 & 0.26 & 0.61 & 0.14 & 0.14 \\ \bottomrule
	\end{tabular}}
%
	\caption{ESTOI improvements for LSTM2 and 5 tested on BBL.}
	\label{tab:stoi_bbl}
	\centering
	\setlength\tabcolsep{5pt} 
	\resizebox{1.0\columnwidth}{!}{%
	\begin{tabular}{c|cccc|cccc}
		\toprule
		& \multicolumn{4}{|c|} {2-Speaker } & \multicolumn{4}{c} {3-Speaker} \\ \midrule
		\begin{tabular}[c]{@{}c@{}}SNR \\ {[dB]}\end{tabular} 	& 
		\begin{tabular}[c]{@{}c@{}}No \\ Proc.\end{tabular} 	& 
		\begin{tabular}[c]{@{}c@{}} IPSF \end{tabular} 			& 
		\begin{tabular}[c]{@{}c@{}}LSTM2\end{tabular} 		& 
		\begin{tabular}[c]{@{}c@{}}LSTM5\end{tabular} 		&   
		\begin{tabular}[c]{@{}c@{}}No \\ Proc.\end{tabular} 	& 
		\begin{tabular}[c]{@{}c@{}} IPSF \end{tabular} 			& 
		\begin{tabular}[c]{@{}c@{}}LSTM2\end{tabular} 		& 
		\begin{tabular}[c]{@{}c@{}}LSTM5\end{tabular} \\ 
		\midrule
					-5 & 0.19 & 0.66 & 0.09 & 0.06 & 0.14 & 0.70 & 0.04 & 0.02 \\ 
			0 & 0.29 & 0.59 & 0.18 & 0.15 & 0.22 & 0.65 & 0.11 & 0.09 \\ 
			5 & 0.39 & 0.51 & 0.21 & 0.20 & 0.29 & 0.60 & 0.15 & 0.14 \\ 
			20 & 0.53 & 0.40 & 0.19 & 0.18 & 0.37 & 0.53 & 0.15 & 0.15 \\ \midrule
			Avg. & 0.35 & 0.54 & 0.17 & 0.15 & 0.26 & 0.62 & 0.11 & 0.10 \\ \bottomrule
	\end{tabular}}
%
	\caption{ESTOI improvements for LSTM3 and 5 tested on STR.}
	\label{tab:stoi_str}
	\centering
	\setlength\tabcolsep{5pt} 
	\resizebox{1.0\columnwidth}{!}{%
	\begin{tabular}{c|cccc|cccc}
		\toprule
		& \multicolumn{4}{|c|} {2-Speaker } & \multicolumn{4}{c} {3-Speaker} \\ \midrule
		\begin{tabular}[c]{@{}c@{}}SNR \\ {[dB]}\end{tabular} 	& 
		\begin{tabular}[c]{@{}c@{}}No \\ Proc.\end{tabular} 	& 
		\begin{tabular}[c]{@{}c@{}} IPSF \end{tabular} 			& 
		\begin{tabular}[c]{@{}c@{}}LSTM3\end{tabular} 		& 
		\begin{tabular}[c]{@{}c@{}}LSTM5\end{tabular} 		&   
		\begin{tabular}[c]{@{}c@{}}No \\ Proc.\end{tabular} 	& 
		\begin{tabular}[c]{@{}c@{}} IPSF \end{tabular} 			& 
		\begin{tabular}[c]{@{}c@{}}LSTM3\end{tabular} 		& 
		\begin{tabular}[c]{@{}c@{}}LSTM5\end{tabular} \\ 
		\midrule
					-5 & 0.24 & 0.60 & 0.16 & 0.15 & 0.18 & 0.65 & 0.10 & 0.09 \\ 
			0 & 0.32 & 0.54 & 0.21 & 0.19 & 0.24 & 0.61 & 0.14 & 0.13 \\ 
			5 & 0.40 & 0.49 & 0.21 & 0.20 & 0.30 & 0.57 & 0.15 & 0.15 \\ 
			20 & 0.54 & 0.39 & 0.18 & 0.18 & 0.37 & 0.53 & 0.15 & 0.15 \\ \midrule
			Avg. & 0.38 & 0.51 & 0.19 & 0.18 & 0.27 & 0.59 & 0.14 & 0.13 \\ \bottomrule
	\end{tabular}}
%
	\caption{ESTOI improvements for LSTM4 and 5 tested on CAF.}
	\label{tab:stoi_caf}
	\centering
	\setlength\tabcolsep{5pt} 
	\resizebox{1.0\columnwidth}{!}{%
	\begin{tabular}{c|cccc|cccc}
		\toprule
		& \multicolumn{4}{|c|} {2-Speaker } & \multicolumn{4}{c} {3-Speaker} \\ \midrule
		\begin{tabular}[c]{@{}c@{}}SNR \\ {[dB]}\end{tabular} 	& 
		\begin{tabular}[c]{@{}c@{}}No \\ Proc.\end{tabular} 	& 
		\begin{tabular}[c]{@{}c@{}} IPSF \end{tabular} 			& 
		\begin{tabular}[c]{@{}c@{}}LSTM4\end{tabular} 		& 
		\begin{tabular}[c]{@{}c@{}}LSTM5\end{tabular} 		&   
		\begin{tabular}[c]{@{}c@{}}No \\ Proc.\end{tabular} 	& 
		\begin{tabular}[c]{@{}c@{}} IPSF \end{tabular} 			& 
		\begin{tabular}[c]{@{}c@{}}LSTM4\end{tabular} 		& 
		\begin{tabular}[c]{@{}c@{}}LSTM5\end{tabular} \\ 
		\midrule
					-5 & 0.24 & 0.60 & 0.13 & 0.12 & 0.19 & 0.65 & 0.08 & 0.07 \\ 
			0 & 0.33 & 0.54 & 0.18 & 0.17 & 0.25 & 0.61 & 0.12 & 0.11 \\ 
			5 & 0.41 & 0.48 & 0.20 & 0.19 & 0.30 & 0.58 & 0.15 & 0.14 \\ 
			20 & 0.53 & 0.39 & 0.18 & 0.18 & 0.37 & 0.53 & 0.15 & 0.15 \\ \midrule
			Avg. & 0.38 & 0.50 & 0.17 & 0.17 & 0.28 & 0.59 & 0.12 & 0.12 \\ \bottomrule
	\end{tabular}}
%
	\caption{ESTOI improvements for LSTM6, 7 and 5 tested on BBL.}
	\label{tab:stoi_bbl_spkr}
	\centering
	\setlength\tabcolsep{5pt} 
	\resizebox{1.0\columnwidth}{!}{%
	\begin{tabular}{c|cccc|cccc}
		\toprule
		& \multicolumn{4}{|c|} {2-Speaker } & \multicolumn{4}{c} {3-Speaker} \\ \midrule
		\begin{tabular}[c]{@{}c@{}}SNR \\ {[dB]}\end{tabular} 	& 
		\begin{tabular}[c]{@{}c@{}}No \\ Proc.\end{tabular} 	& 
		\begin{tabular}[c]{@{}c@{}} IPSF \end{tabular} 			& 
		\begin{tabular}[c]{@{}c@{}}LSTM6\end{tabular} 		& 
		\begin{tabular}[c]{@{}c@{}}LSTM5\end{tabular} 		&   
		\begin{tabular}[c]{@{}c@{}}No \\ Proc.\end{tabular} 	& 
		\begin{tabular}[c]{@{}c@{}} IPSF \end{tabular} 			& 
		\begin{tabular}[c]{@{}c@{}}LSTM7\end{tabular} 		& 
		\begin{tabular}[c]{@{}c@{}}LSTM5\end{tabular} \\ 
		\midrule
					-5 & 0.20 & 0.66 & 0.07 & 0.06 & 0.14 & 0.69 & 0.02 & 0.02 \\ 
			0 & 0.30 & 0.59 & 0.16 & 0.16 & 0.22 & 0.65 & 0.08 & 0.09 \\ 
			5 & 0.39 & 0.52 & 0.20 & 0.20 & 0.29 & 0.60 & 0.13 & 0.14 \\ 
			20 & 0.54 & 0.40 & 0.17 & 0.19 & 0.38 & 0.53 & 0.14 & 0.15 \\ \midrule
			Avg. & 0.36 & 0.54 & 0.15 & 0.15 & 0.26 & 0.62 & 0.09 & 0.10 \\ \bottomrule
	\end{tabular}}
%
%
	\caption{ESTOI improvements for LSTM5 tested on BUS $|$ PED.}
	\label{tab:stoi_ped}
	\centering
	\setlength\tabcolsep{5pt} 
	\resizebox{1.0\columnwidth}{!}{%
	\begin{tabular}{c|ccc|ccc}
		\toprule
		& \multicolumn{3}{|c|} {2-Speaker } & \multicolumn{3}{c} {3-Speaker} \\ \midrule
		\begin{tabular}[c]{@{}c@{}}SNR \\ {[dB]}\end{tabular} 	& 
		\begin{tabular}[c]{@{}c@{}}No \\ Proc.\end{tabular} 	& 
		\begin{tabular}[c]{@{}c@{}} IPSF \end{tabular} 			& 
		\begin{tabular}[c]{@{}c@{}}LSTM5\end{tabular} 		&   
		\begin{tabular}[c]{@{}c@{}}No \\ Proc.\end{tabular} 	& 
		\begin{tabular}[c]{@{}c@{}} IPSF \end{tabular} 			& 
		\begin{tabular}[c]{@{}c@{}}LSTM5\end{tabular} \\ 
		\midrule
					-5 & 0.32 $|$ 0.18 & 0.55 $|$ 0.64 & 0.14 $|$ 0.08 & 0.24 $|$ 0.14 & 0.61 $|$ 0.68 & 0.08 $|$ 0.03 \\ 
			0 & 0.39 $|$ 0.28 & 0.50 $|$ 0.58 & 0.18 $|$ 0.15 & 0.28 $|$ 0.21 & 0.58 $|$ 0.63 & 0.12 $|$ 0.09 \\ 
			5 & 0.45 $|$ 0.37 & 0.46 $|$ 0.52 & 0.20 $|$ 0.20 & 0.32 $|$ 0.28 & 0.56 $|$ 0.59 & 0.14 $|$ 0.13 \\ 
			20 & 0.55 $|$ 0.54 & 0.39 $|$ 0.40 & 0.18 $|$ 0.18 & 0.38 $|$ 0.37 & 0.53 $|$ 0.53 & 0.15 $|$ 0.15 \\ \midrule
			Avg. & 0.43 $|$ 0.34 & 0.47 $|$ 0.54 & 0.17 $|$ 0.15 & 0.31 $|$ 0.25 & 0.57 $|$ 0.61 & 0.12 $|$ 0.10 \\ \bottomrule
	\end{tabular}}
\end{table}

\section{Conclusion}\label{sec:conclusion}
In this paper we have proposed utterance-level Permutation Invariant Training\;(uPIT) for speaker independent multi-talker speech separation and denoising.
Differently from prior works, that focus only on the ideal noise-free setting, we focus on the more realistic scenario of speech separation in noisy environments.  
Specifically, using the uPIT technique we have trained bi-directional Long Short-Term Memory\;(LSTM) Recurrent Neural Networks\;(RNNs), to separate two and three-speaker mixtures corrupted by multiple noise types at a wide range of Signal to Noise Ratios\;(SNRs).       

We show that bi-directional LSTM RNNs trained with uPIT are capable of improving both Signal to Distortion Ratio\;(SDR), as well as the Extended Short-Time Objective Intelligibility\;(ESTOI) measure for challenging noise types and SNRs. 
Specifically, we show that LSTM RNNs achieve large SDR and ESTOI improvements, when evaluated using noise types seen during training, and that a single model is capable of handling multiple noise types with only a slight decrease in performance. 
Furthermore, we show that a single LSTM RNN can handle both two-speaker and three-speaker noisy mixtures, without \emph{a priori} knowledge about the exact number of speakers. 
Finally, we show that LSTM RNNs trained using uPIT generalizes well to unknown noise types.   



\bibliographystyle{IEEEtran}
\bibliography{mybib}

\begin{thebibliography}{10}
\providecommand{\url}[1]{#1}
\csname url@samestyle\endcsname
\providecommand{\newblock}{\relax}
\providecommand{\bibinfo}[2]{#2}
\providecommand{\BIBentrySTDinterwordspacing}{\spaceskip=0pt\relax}
\providecommand{\BIBentryALTinterwordstretchfactor}{4}
\providecommand{\BIBentryALTinterwordspacing}{\spaceskip=\fontdimen2\font plus
\BIBentryALTinterwordstretchfactor\fontdimen3\font minus
  \fontdimen4\font\relax}
\providecommand{\BIBforeignlanguage}[2]{{%
\expandafter\ifx\csname l@#1\endcsname\relax
\typeout{** WARNING: IEEEtran.bst: No hyphenation pattern has been}%
\typeout{** loaded for the language `#1'. Using the pattern for}%
\typeout{** the default language instead.}%
\else
\language=\csname l@#1\endcsname
\fi
#2}}
\providecommand{\BIBdecl}{\relax}
\BIBdecl

\bibitem{bronkhorst_cocktail_2000}
A.~W. Bronkhorst, ``The {Cocktail} {Party} {Phenomenon}: {A} {Review} of
  {Research} on {Speech} {Intelligibility} in {Multiple}-{Talker}
  {Conditions},'' \emph{Acta Acust united Ac}, vol.~86, no.~1, pp. 117--128,
  2000.

\bibitem{mcdermott_cocktail_2009}
J.~H. McDermott, ``\BIBforeignlanguage{English}{The cocktail party problem},''
  \emph{\BIBforeignlanguage{English}{Current Biology}}, vol.~19, no.~22, pp.
  R1024--R1027, Dec. 2009.

\bibitem{kolbaek_speech_2017}
M.~Kolbæk, Z.~H. Tan, and J.~Jensen, ``Speech {Intelligibility} {Potential} of
  {General} and {Specialized} {Deep} {Neural} {Network} {Based} {Speech}
  {Enhancement} {Systems},'' \emph{IEEE/ACM Trans. Audio, Speech, Lang.
  Process.}, vol.~25, no.~1, pp. 153--167, 2017.

\bibitem{chen_long_2016}
J.~Chen and D.~Wang, ``Long {Short}-{Term} {Memory} for {Speaker}
  {Generalization} in {Supervised} {Speech} {Separation},'' in \emph{Proc.
  {INTERSPEECH}}, 2016, pp. 3314 -- 3318.

\bibitem{weninger_single-channel_2014}
F.~Weninger, F.~Eyben, and B.~Schuller, ``Single-channel speech separation with
  memory-enhanced recurrent neural networks,'' in \emph{Proc. {ICASSP}}, 2014,
  pp. 3709--3713.

\bibitem{weninger_discriminatively_2014}
F.~Weninger \emph{et~al.}, ``Discriminatively trained recurrent neural networks
  for single-channel speech separation,'' in \emph{{GlobalSIP}}, 2014, pp.
  577--581.

\bibitem{erdogan_phase-sensitive_2015}
H.~Erdogan \emph{et~al.}, ``Phase-sensitive and recognition-boosted speech
  separation using deep recurrent neural networks,'' in \emph{Proc. {ICASSP}},
  2015, pp. 708--712.

\bibitem{chen_large-scale_2016}
J.~Chen \emph{et~al.}, ``Large-scale training to increase speech
  intelligibility for hearing-impaired listeners in novel noises,'' \emph{J.
  Acoust. Soc. Am.}, vol. 139, no.~5, pp. 2604--2612, 2016.

\bibitem{du_speech_2014}
J.~Du \emph{et~al.}, ``Speech separation of a target speaker based on deep
  neural networks,'' in \emph{{ICSP}}, 2014, pp. 473--477.

\bibitem{yu_permutation_2017}
D.~Yu \emph{et~al.}, ``Permutation {Invariant} {Training} of {Deep} {Models}
  for {Speaker}-{Independent} {Multi}-talker {Speech} {Separation},'' in
  \emph{Proc. {ICASSP}}, 2017, pp. 241--245.

\bibitem{hershey_deep_2016}
J.~R. Hershey \emph{et~al.}, ``Deep clustering: {Discriminative} embeddings for
  segmentation and separation,'' in \emph{Proc. {ICASSP}}, 2016, pp. 31--35.

\bibitem{isik_single-channel_2016}
Y.~Isik \emph{et~al.}, ``Single-{Channel} {Multi}-{Speaker} {Separation}
  {Using} {Deep} {Clustering},'' in \emph{Proc. {INTERSPEECH}}, 2016, pp.
  545--549.

\bibitem{chen_deep_2017}
Z.~Chen, Y.~Luo, and N.~Mesgarani, ``Deep attractor network for
  single-microphone speaker separation,'' in \emph{Proc. {ICASSP}}, 2017, pp.
  246--250.

\bibitem{weng_deep_2015}
C.~Weng \emph{et~al.}, ``Deep {Neural} {Networks} for {Single}-{Channel}
  {Multi}-{Talker} {Speech} {Recognition},'' \emph{IEEE/ACM Trans. Audio,
  Speech, Lang. Process.}, vol.~23, no.~10, pp. 1670--1679, 2015.

\bibitem{huang_joint_2015}
P.-S. Huang \emph{et~al.}, ``Joint {Optimization} of {Masks} and {Deep}
  {Recurrent} {Neural} {Networks} for {Monaural} {Source} {Separation},''
  \emph{IEEE/ACM Trans. Audio, Speech, Lang. Process.}, vol.~23, no.~12, pp.
  2136--2147, 2015.

\bibitem{goodfellow_deep_2016}
I.~Goodfellow, Y.~Bengio, and A.~Courville, \emph{Deep {Learning}}.\hskip 1em
  plus 0.5em minus 0.4em\relax MIT Press, 2016.

\bibitem{gulshan_development_2016}
V.~Gulshan \emph{et~al.}, ``Development and {Validation} of a {Deep} {Learning}
  {Algorithm} for {Detection} of {Diabetic} {Retinopathy} in {Retinal} {Fundus}
  {Photographs},'' \emph{JAMA}, vol. 316, no.~22, pp. 2402--2410, 2016.

\bibitem{esteva_dermatologist-level_2017}
A.~Esteva \emph{et~al.}, ``\BIBforeignlanguage{en}{Dermatologist-level
  classification of skin cancer with deep neural networks},''
  \emph{\BIBforeignlanguage{en}{Nature}}, vol. 542, no. 7639, pp. 115--118,
  2017.

\bibitem{xiong_achieving_2016}
W.~Xiong \emph{et~al.}, ``Achieving {Human} {Parity} in {Conversational}
  {Speech} {Recognition},'' \emph{arXiv:1610.05256 [cs]}, 2016.

\bibitem{saon_english_2017}
G.~Saon \emph{et~al.}, ``English {Conversational} {Telephone} {Speech}
  {Recognition} by {Humans} and {Machines},'' \emph{arXiv:1703.02136 [cs]},
  2017.

\bibitem{healy_algorithm_2015}
E.~W. Healy \emph{et~al.}, ``An algorithm to increase speech intelligibility
  for hearing-impaired listeners in novel segments of the same noise type,''
  \emph{J. Acoust. Soc. Am.}, vol. 138, no.~3, pp. 1660--1669, 2015.

\bibitem{goehring_speech_2017}
T.~Goehring \emph{et~al.}, ``Speech enhancement based on neural networks
  improves speech intelligibility in noise for cochlear implant users,''
  \emph{Hearing Research}, vol. 344, pp. 183--194, 2017.

\bibitem{erdogan_deep_2017}
H.~Erdogan \emph{et~al.}, ``Deep {Recurrent} {Networks} for {Separation} and
  {Recognition} of {Single} {Channel} {Speech} in {Non}-stationary {Background}
  {Audio},'' in \emph{New {Era} for {Robust} {Speech} {Recognition}:
  {Exploiting} {Deep} {Learning}}.\hskip 1em plus 0.5em minus 0.4em\relax
  Springer, 2017.

\bibitem{kolbaek_multi-talker_2017-1}
M.~Kolbæk \emph{et~al.}, ``Multi-talker {Speech} {Separation} with
  {Utterance}-level {Permutation} {Invariant} {Training} of {Deep} {Recurrent}
  {Neural} {Networks},'' \emph{IEEE/ACM Trans. Audio, Speech, Lang. Process.},
  vol.~25, no.~10, pp. 1901--1913, 2017.

\bibitem{wang_training_2014}
Y.~Wang, A.~Narayanan, and D.~Wang, ``On {Training} {Targets} for {Supervised}
  {Speech} {Separation},'' \emph{IEEE/ACM Trans. Audio, Speech, Lang.
  Process.}, vol.~22, no.~12, pp. 1849--1858, 2014.

\bibitem{garofolo_csr-i_1993}
J.~Garofolo \emph{et~al.}, ``{CSR}-{I} ({WSJ}0) {Complete} {LDC}93s6a,'' 1993,
  philadelphia: Linguistic Data Consortium.

\bibitem{noauthor_itu_1993}
``{ITU} {Rec}. {P}.56 : {Objective} measurement of active speech level,''
  1993, https://www.itu.int/rec/T-REC-P.56/.

\bibitem{garofolo_darpa_1993}
J.~S. Garofolo \emph{et~al.}, ``{DARPA} {TIMIT} {Acoustic} {Phonetic}
  {Continuous} {Speech} {Corpus} {CDROM},'' 1993.

\bibitem{barker_third_2015}
J.~Barker \emph{et~al.}, ``The third '{CHiME}' {Speech} {Separation} and
  {Recognition} {Challenge}: {Dataset}, task and baselines,'' in \emph{Proc.
  {ASRU}}, 2015.

\bibitem{hochreiter_long_1997}
S.~Hochreiter and J.~Schmidhuber, ``\BIBforeignlanguage{eng}{Long short-term
  memory},'' \emph{\BIBforeignlanguage{eng}{Neural Comput}}, vol.~9, no.~8, pp.
  1735--1780, 1997.

\bibitem{agarwal_introduction_2014}
A.~Agarwal \emph{et~al.}, ``An introduction to computational networks and the
  computational network toolkit,'' Microsoft Technical Report
  \{MSR-TR\}-2014-112, Tech. Rep., 2014.

\bibitem{vincent_performance_2006}
E.~Vincent, R.~Gribonval, and C.~Fevotte, ``Performance measurement in blind
  audio source separation,'' \emph{IEEE/ACM Trans. Audio, Speech, Lang.
  Process.}, vol.~14, no.~4, pp. 1462--1469, 2006.

\bibitem{jensen_algorithm_2016}
J.~Jensen and C.~H. Taal, ``An {Algorithm} for {Predicting} the
  {Intelligibility} of {Speech} {Masked} by {Modulated} {Noise} {Maskers},''
  \emph{IEEE/ACM Trans. Audio, Speech, Lang. Process.}, vol.~24, no.~11, pp.
  2009--2022, 2016.

\bibitem{loizou_speech_2013}
P.~C. Loizou, \emph{Speech {Enhancement}: {Theory} and {Practice}}.\hskip 1em
  plus 0.5em minus 0.4em\relax CRC Press, 2013.

\bibitem{erkelens_minimum_2007}
J.~S. Erkelens \emph{et~al.}, ``Minimum {Mean}-{Square} {Error} {Estimation} of
  {Discrete} {Fourier} {Coefficients} {With} {Generalized} {Gamma} {Priors},''
  \emph{IEEE/ACM Trans. Audio, Speech, Lang. Process.}, vol.~15, no.~6, pp.
  1741--1752, 2007.

\end{thebibliography}

\end{document}